\documentclass{nature}
\usepackage{graphicx}
\usepackage{amsmath}

\newcommand{\be}{\begin{equation}}
\newcommand{\ee}{\end{equation}}

\title{High-throughput Imaging of Self-luminous Objects through a Single Optical Fiber}

\author{Roman Barankov$^{1}$ \& Jerome Mertz$^{1}$}

\begin{document}

\maketitle

\begin{affiliations}
 \item Department of Biomedical Engineering, Boston University, 44 Cummington Mall,  Boston, MA 02215, USA
\end{affiliations}

\begin{abstract}
Imaging through a single optical fiber offers attractive possibilities in many applications such as microendoscopy or remote sensing. However, the direct transmission of an image through an optical fiber is difficult because spatial information is scrambled upon propagation. We demonstrate an image transmission strategy where spatial information is first converted to spectral information. Our strategy is based on a principle of spread-spectrum encoding, borrowed from wireless communications, wherein object pixels are converted into distinct spectral codes that span the full bandwidth of the object spectrum. Image recovery is performed by numerical inversion of the detected spectrum at the fiber output. We provide a simple demonstration of spread-spectrum encoding using Fabry-Perot etalons. Our technique enables the 2D imaging of self-luminous (i.e. incoherent) objects with high throughput in principle independent of pixel number. Moreover, it is insensitive to fiber bending, contains no moving parts, and opens the possibility of extreme miniaturization.
\end{abstract}

Optical imaging devices have become ubiquitous in our society, and a trend
toward their miniaturization has been inexorable. In addition to
facilitating portability, miniaturization can enable imaging of targets that
are difficult to access. For example, in the biomedical field, miniaturized
endoscopes can provide microscopic images of subsurface structures within
tissue \cite{schnitzer}. Such imaging is generally based on the use of miniaturized lenses,
though for extreme miniaturization lensless strategies may be required \cite{gill}.

Particularly attractive is the possibility of imaging through a single, bare optical fiber. The transmission of spatial information through an optical fiber can be
achieved in many ways \cite{yariv,friesem}. One example is by modal multiplexing
where different spatial distributions of light are coupled to different
spatial modes of a multimode fiber. Propagation through the fiber scrambles
these modes, but these can be unscrambled if the transmission matrix \cite{mosk, gigan} of the
fiber is measured a priori \cite{dileonardo,dhokalia, choi,psaltis,piestun,kahn}. Such spatio-spatial encoding can lead to high
information capacity \cite{stuart} but suffers from the problem that the
transmission matrix is not robust. Any motion or bending of the fiber
requires a full recalibration of this matrix \cite{psaltis,piestun}, which is problematic
while imaging. A promising alternative is spatio-spectral encoding,
since the spectrum of light propagating through a fiber is relatively
insensitive to fiber motion or bending. Moreover, such encoding presumes
that the light incident on the fiber is spectrally diverse, or broadband,
which is fully compatible with our goal here of imaging self-luminous
sources.

Techniques already exist to convert spatial information into spectral
information. For example, a prism or grating maps different directions of a
light rays into different colors. By placing a miniature grating and lens in
front of an optical fiber, directional (spatial) information can be
converted into color (spectral) information, and launched into the fiber.
Such a technique has been used to perform 1D imaging of transmitting or
reflecting \cite{friesem, kartashev, bartelt, tearney} or even self-luminous
\cite{tai,yelin} objects, where 2D imaging is then obtained by a mechanism of
physical scanning along the orthogonal axis. Alternatively, scanningless 2D
imaging with no moving parts has been performed by angle-wavelength encoding
\cite{friesem} or fully spectral encoding using a combination of gratings
\cite{lohmann} or a grating and a virtual image phased array (VIPA) \cite{weiner,jalali}. These 2D techniques have only been applied to non-self-luminous objects. Such techniques of spectral encoding using a grating have been implemented in clinical endoscope
configurations only recently. To our knowledge, the smallest diameter of
such an endoscope is $350\,\rm\mu m$, partly limited by the requirement of a miniature
lens in the endoscope \cite{tearney2006}.

A property of prisms or gratings is that they spread different colors into
different directions. For example, in the case of spectrally encoded
endoscopy \cite{tearney}, broadband light from a fiber is collimated by a lens and then
spread by a grating into many rays traveling in different directions. Such a
spreading of directions defines the field of view of the endoscope. If we consider the inverse problem of detecting broadband light from a self-luminous source using the same grating/lens
configuration, we find that because of this property of directional spreading
only a fraction of the spectral power from the source is channeled by
the grating into the fiber. The rest of the power physically misses the
fiber entrance, which plays the role of a spectral slit, and becomes
lost. Indeed, if the field of view is divided into $M$ resolvable
directions (object pixels), then at best only the fraction $1/M$ of the
power from each pixel can be detected. Such a scaling law is
inefficient and prevents the scaling of this spectral encoding mechanism to
many pixels (a similar problem occurs when using a randomly scattering medium instead of a grating \cite{dogariu,cao}). Given that the power from self-luminous sources is generally
limited, then certainly it is desirable to not throw away most of this power.

Our solution to this problem
involves the use of a spectral encoder that does
not spread the direction of an incoming ray, imprints a code onto the ray
spectrum depending on the ray direction, and this code occupies the full
bandwidth of the spectrum. As a result of these properties, the fraction of
power that can be detected from any given object pixel is roughly fixed and does
not decrease as $1/M$. We call our encoder a spread-spectrum encoder (SSE).
Indeed, there is a close analogy with strategies used in wireless
communication  \cite{sklar}. While the spectral encoding techniques described above that
involve gratings perform the equivalent of wavelength-division multiplexing,
our technique performs the equivalent of code-division multiplexing (or
CDMA).

\section*{Results}

\subsection{Design of a spread-spectrum encoder\\}

\begin{figure}
\centerline{\includegraphics[width=12cm]{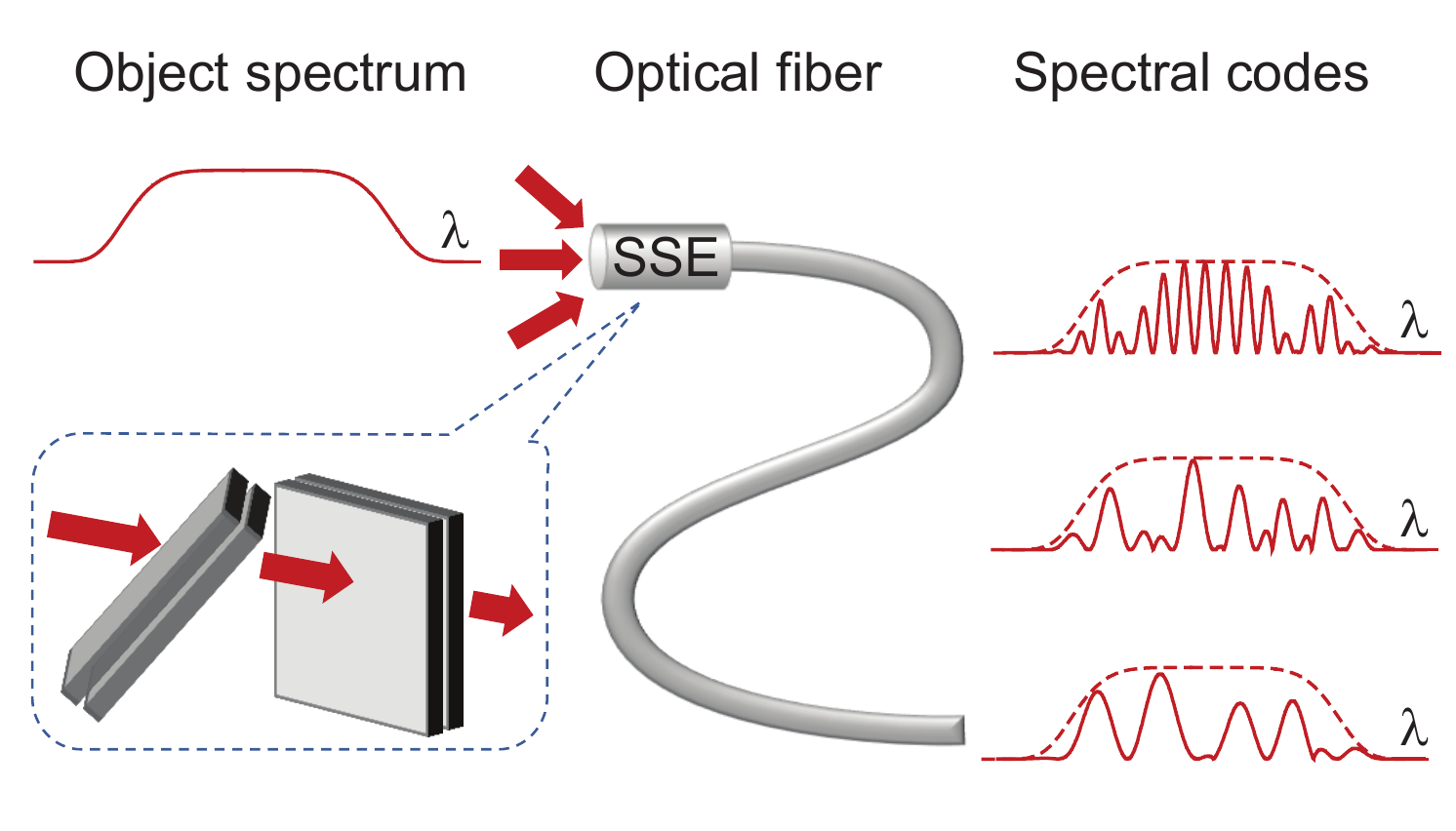}}
\caption{{\bf Schematic of method.}
A self-luminous object with an extended spatial distribution produces light with a broad spectrum. Different spatial portions (pixels) of this object are incident on the entrance of an optical fiber from different directions. A spread-spectrum encoder (SSE) imparts a unique spectral code to each light direction in a power efficient manner. The resulting total spectrum at the output of the fiber is detected by a spectrograph, and image reconstruction is performed by numerical decoding. An example of a SSE is shown in the inset, consisting of two low finesse Fabry-Perot etalons of different free-spectral ranges, tilted perpendicular to each other to encode light directions in 2D.}
\label{Schematic}
\end{figure}

A key property of a SSE is that it should spread the direction of incoming rays as little as possible. To ensure this, we must first identify where such spreading comes from. In the case of a grating it comes from lateral
features in the grating structure. A SSE, therefore, should be essentially devoid of
lateral features and as translationally invariant as possible. On the
other hand, to impart spectral codes, it must produce wavelength-dependent
time delays. An example of a SSE that satisfies these conditions is a low
finesse Fabry-Perot etalon (FPE), as illustrated in the inset of Fig.~\ref{Schematic}. This is translationally invariant and
thus does not alter ray directions. Moreover, different wavelengths travelling through
the etalon experience different, multiple time delays owing to multiple
reflections. These time delays also depend on ray direction, providing the
possibility of angle-wavelength encoding. It is important to note that
losses through such a device are in the axial (backward) direction only, as
opposed to lateral directions. Hence, there is no slit effect as in the case
of a grating, meaning that the full spectrum can be utilized for encoding,
and losses can be kept to a minimum independent of pixel number. This advantage is
akin to the Jacquinot advantage in interferometric spectroscopy \cite{jacquinot} (also called the \'etendue advantage).

\begin{figure}
\centerline{\includegraphics[width=12cm]{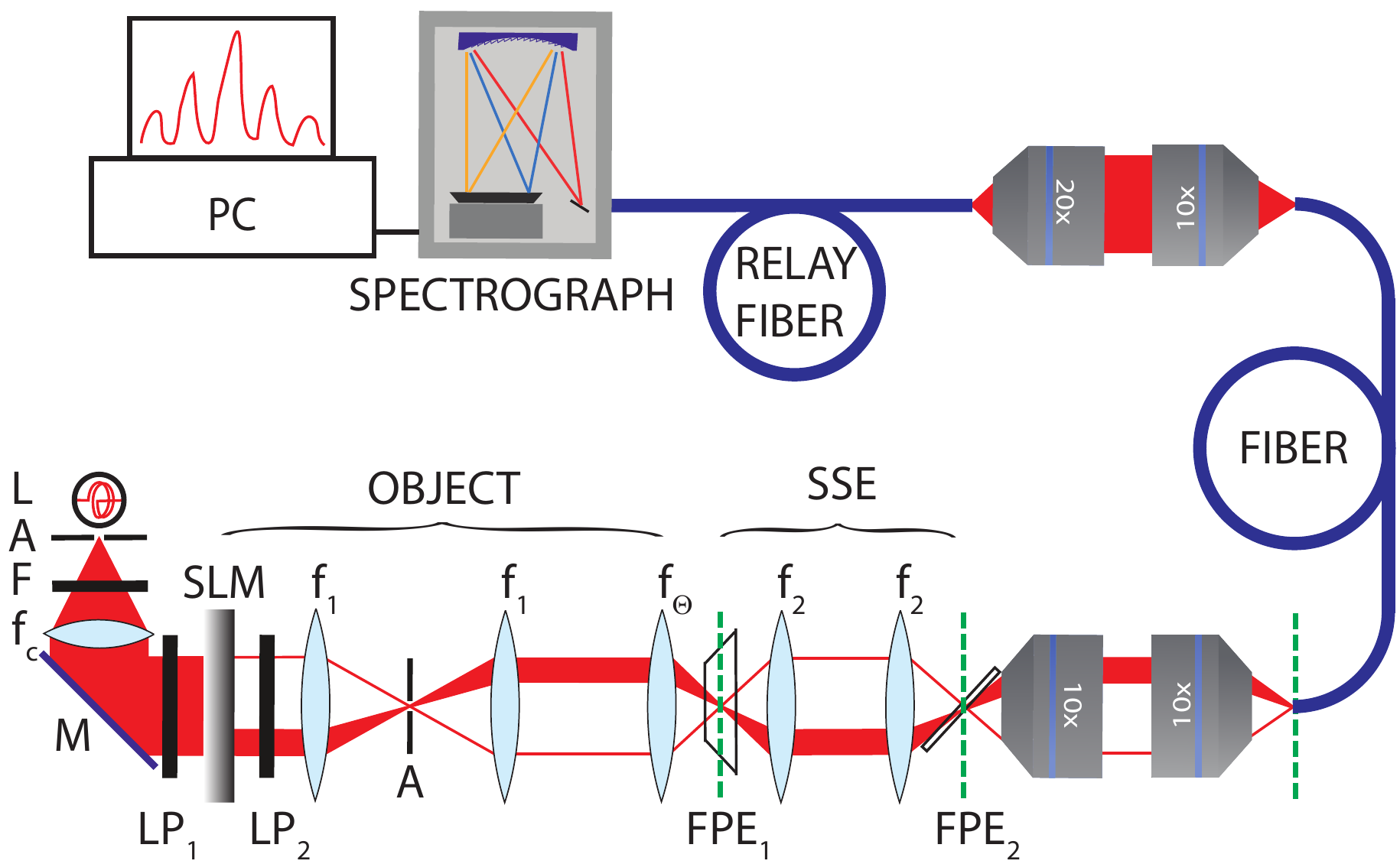}}
\caption{{\bf Experimental setup for demonstration of spread-spectrum encoding.}
 White light from a lamp is collimated with the use of a $1 \,\rm mm$ aperture (A) and collimating lens ($\textrm{f}_c=100\,\rm mm$). The light is directed through a transmission spatial light modulator (SLM). Crossed linear polarizers ($\textrm{LP}_1$ and $\textrm{LP}_2$) ensure that the SLM provides amplitude modulation. A short-pass filter (F) is inserted to block wavelengths longer than $740\,\rm nm$ where the SLM contrast becomes degraded. A unit magnification relay ($\textrm{f}_1=100\,\rm mm$) enables the insertion of a second $1 \,\rm mm$ aperture (A) to remove spurious high order diffracted light from the SLM. Lens ($\textrm{f}_\theta=25\,\rm mm$) plays a key role in converting SLM pixel positions into ray directions. The SSE is comprised of two orthogonally tilted low finesse Fabry-Perot etalons ($\textrm{FPE}_1$ and $\textrm{FPE}_2$), separated by a unit magnification relay ($\textrm{f}_2=25\,\rm mm$). The light exiting the SSE is relayed ($10\times$, 0.25NA objectives) to the entrance of a bare optical fiber of core diameter $200 \,\rm \mu m$. A secondary relay fiber with smaller core size ($100 \,\rm \mu m$) serves to improve the spectral resolution of our spectrograph. The spectrum is recorded with a camera and sent to a computer (PC). The vertical dashed lines correspond to optical planes separated by unit magnification relays whose only purpose is to provide physical space to accommodate optical mounts.}
\label{SSE_setup}
\end{figure}

A more detailed schematic of our setup is shown in Fig. \ref{SSE_setup} and described in
Methods. To simulate an arbitrary angular distribution of self-luminous
sources, we made use of a spatial light modulator (SLM) and a lens. The SLM
was trans-illuminated by white light from a lamp and the lens's only purpose
was to convert spatial coordinates at the SLM to angular coordinates at the
SSE entrance. The SSE here consisted of two FPEs in tandem, each tilted off axis so
as to eliminate angular encoding degeneracies about the main optical axis
(note: such tilting still preserves translational invariance). In this
manner, each 2D ray direction within an angle spanning about 160 mrad, corresponding to our field of view, was encoded
into a unique spectral pattern which was then launched into a multimode
fiber of larger numerical aperture. The light spectrum at the proximal end of
the fiber was then measured with a spectrograph and sent to a computer for
interpretation.

\subsection{Image retrieval\\}

In our lensless configuration, the optical angular resolution defined by the fiber core diameter was about 3 mrad, which was better (smaller) than the minimum angular pixel size (12 mrad) used to encode our objects. Moreover, these minimum object pixel sizes were large enough to be spatially incoherent, meaning that the spectral signals produced by the pixels
could be considered as independent of one another, thus reducing SSE to a
linear system. Specifically, for $M$ input pixel elements (ray directions $%
A_{m}$), and $N$ output spectral detection elements ($B_{n}$), the SSE
process can be written in matrix form as $\rm{B=\mathbf{M}A}$, where each column
in $\mathbf{M}$ corresponds to the spectral code for its associated object pixel.
These spectral codes must be measured in advance prior to any imaging experiment; however,
once measured, they are insensitive to fiber
motion or bending. Once $\mathbf{M}$ is defined, then imaging can be performed. An arbitrary distribution of self-luminous sources (object pixels) leads to a
superposition of spectral codes weighted according to their respective pixel
intensities. A retrieval of this distribution ($\rm{A}$) from a measurement of
the total output spectrum ($\rm{B}$), is then formally given by $\rm{A=\mathbf{M}^{+}B}$,
where $\mathbf{M}^{+}$ is the pseudo-inverse of $\mathbf{M}$ (see Methods).

\begin{figure}
\centerline{\includegraphics[width=12cm]{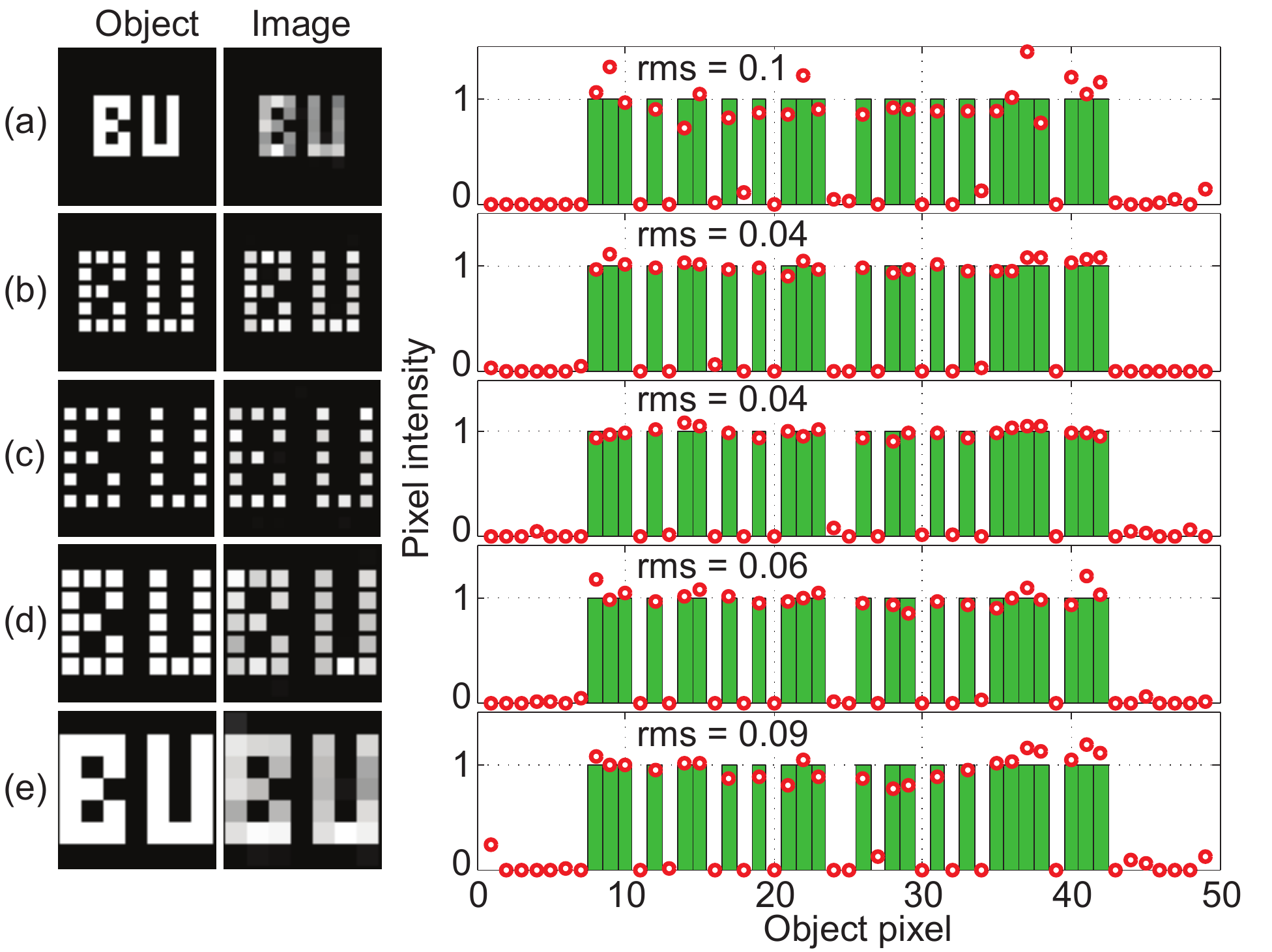}}
\caption{{\bf Influence of pixel size and sparcity on SSE imaging.} Spatially incoherent, white-light objects in the shape of ``BU", consisting of 49 pixels ($7\times7$), are sent through a SSE and launched into a fiber (see Methods). The reconstructed images and pixel values, along with associated rms errors, are displayed for different pixel layouts. In rows (a)-(c), the pixel sizes (powers) are the same, but their separation is increased, leading to better reconstruction (lower rms error). In rows (c)-(e) the pixel separations are the same, but their size is increased until the pixels are juxtaposed. The corresponding increase in pixel power is counterbalanced by a decrease in the contrast of associated spectral codes (due to increased angular diversity per pixel), leading to a net decrease in reconstruction quality (higher rms). The reconstructions were verified to be stable and insensitive to fiber movement or bending (see Supplementary Figure 3).}
\label{BU}
\end{figure}

\begin{figure}
\centerline{\includegraphics[width=12cm]{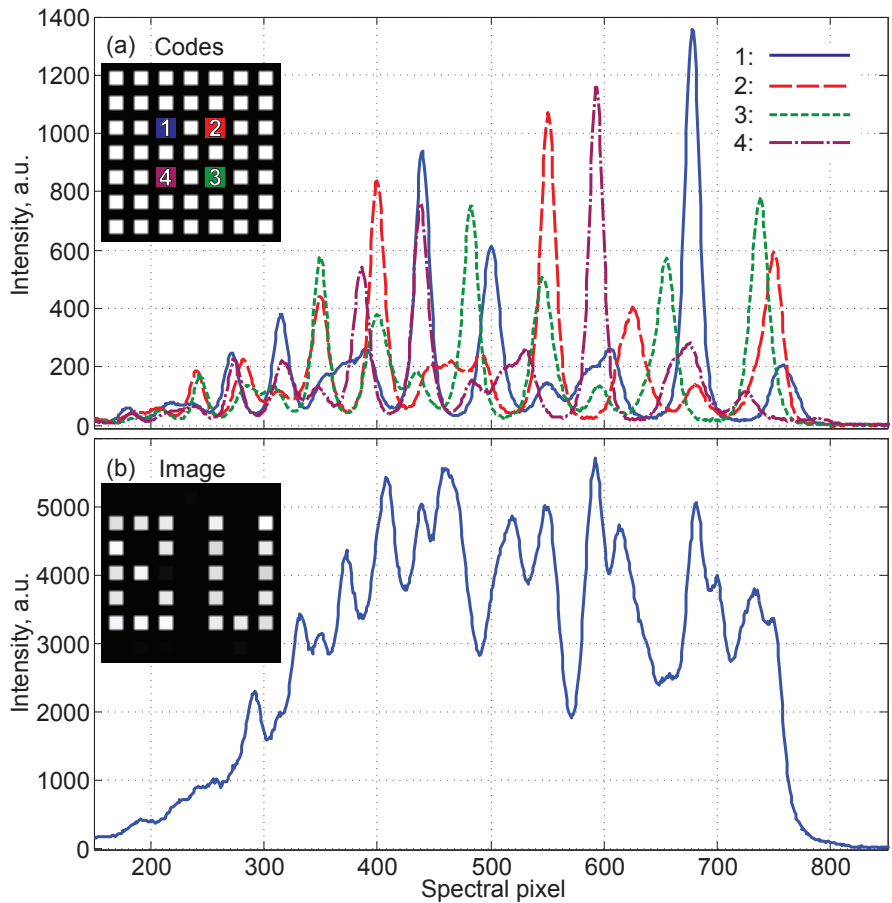}}
\caption{{\bf Examples of spectral codes and detected spectrum.} Spectral codes and output spectrum associated with Fig. 2c. (a) Representative spectral codes (columns of \textbf{M}) for four object pixels near center (highlighted by different colors in the inset). (b) The total output spectrum (B) resulting from the weighted sum of spectra from all object pixels, here in the shape of ``BU" shown in the inset. One spectral pixel spans a wavelength of about $0.5\,\rm nm$, leading to a full object bandwidth of about $300\,\rm nm$. The spectral pixel marked $500$ corresponds to $565\,\rm nm$.}
\label{SVD_codes_BU_spectra}
\end{figure}

While the high throughput of a SSE facilitates the efficient transmission of object brightness, this alone is not sufficient to guaranty that image retrieval will be reliable.  As we will see below, the reliability and immunity to noise of image retrieval critically depends on the
condition of $\mathbf{M}$. For example, representative spectral codes and singular values of $\mathbf{M}$ obtained with our SSE are
illustrated in Supplementary Figure 1. Ideally, the singular values should be as evenly
distributed as possible, indicating highly orthogonal spectral codes. In our
case, condition numbers of $\mathbf{M}$ were on the order of a few hundred,
suggesting that our SSEs were not ideal and could be significantly improved
in future designs. Nevertheless, despite these high condition numbers, we were able to retrieve 2D images up to $49$ pixels in size (see Figs. \ref{BU},\ref{SVD_codes_BU_spectra}, and Supplementary Figure 2), with weak pixel brightnesses roughly corresponding to that of a dimly lit room.

\subsection{Number of resolvable pixels\\}

To evaluate the maximum number of resolvable object pixels that can be encoded by our SSE, we consider the idealized case where
shot-noise is the only noise source and begin by estimating the signal to
noise ratio (SNR) for the image retrieval of a single pixel. 
We define $W$ to be the total power that would be detected if no SSE were
present in the system and all object pixels were full on throughout the
entire field of view. We further
define $\eta $ to be the average transmissivity of the SSE (dependent on throughput), and $\gamma$ to
be a measure of the information content in the spectral codes defined by the coding matrix $\mathbf{M}$ (largely independent of throughput). This last parameter is difficult to quantify as it depends on the quality of the spectral codes, and thereby on the number of object pixels $M$ enclosed in the field of view. Clearly, if the
spectral codes are completely redundant ($\gamma=0$) then no image
information can be transmitted. A lower limit to $\gamma$ can be roughly estimated by the inverse condition number of the coding matrix $\mathbf{M}$~\cite{Golub_Loan}. However, this lower limit can be significantly increased by applying dimensional reduction to $\mathbf{M}$ (i.e. applying a minimum threshold to the singular values of $\mathbf{M}$ upon pseudoinversion) or by applying priors during image reconstruction, such as a positivity constraint. In our case, with typical separations and sizes of $M=49$ object pixels and with the application of a positivity constraint, we attain values of $\gamma$ on the order of $0.01$ (though often a bit smaller). This is far short of the maximum value $\gamma=1$ which could be attained with ideal codes, and is an indication that there remains considerable room for improvement in the design of better SSE's.      

Continuing with our calculation, if the field of view is subdivided into $M$ object pixels, then the detected
energy per object pixel after passing through the SSE is $\eta WT/M$, where $T$ is the measurement time. In accord with the principle of spread-spectrum encoding, this
energy is distributed roughly uniformly over $N$ detector (spectrograph) pixels, meaning that
the average energy per detector pixel is $\eta WT/MN$, and the corresponding
rms noise per detector pixel is $\sqrt{\eta WT/MN}$, where we assume $W$ is
measured in units of photoelectrons/s and we neglect detector noise. But not all this
average energy is useful. The energy per detection pixel that actually
contains image information is $\gamma \eta WT/MN$. We thus have a measure of
signal and noise at the spectrograph.

To calculate the signal and noise after image retrieval, we make the
assumption that all the detector signals are correlated and thus
contribute to the retrieved image signal in a coherent way (i.e. scaling
with $N$), whereas all the detector noises are uncorrelated and contribute
to the retrieved image noise in an incoherent way (i.e. scaling with $\sqrt{N}$). The $\text{SNR}$ of the retrieved image per object pixel, assuming $N>M$, is thus

\be
\text{SNR}_A\simeq\gamma\,\text{SNR}_B=\gamma\sqrt{\eta \frac{WT}{M}}, 
\ee
where $\text{SNR}_B$ is the SNR associated with the detected spectrum $\text{B}$ per object pixel (the first relation is in accord with Ref.~\cite{Golub_Loan}). 

This expression warrants scrutiny. First, let us examine how $\text{SNR}_A$ depends on 
$M$. As $M$ increases, then the power per object pixel $W/M$ decreases, which
is obviously detrimental to $\text{SNR}_A$. On the other hand, as $M$ decreases, then, provided the field of view remains fixed, the object pixels become larger in size, meaning they become
spread over larger solid angles. This increase in angular diversity per
pixel generally leads to a decrease in the contrast $\gamma$ of the spectral codes
associated with each pixel. That is, both large and small $M$'s are
ultimately detrimental to $\text{SNR}_A$.

At first glance it might appear
that $\text{SNR}_A$ does not depend on $N$, until we note that $\gamma$ depends on the contrast of the spectral codes upon detection. If $N$ is large and
the spectral codes are oversampled upon detection, then $N$ has no effect on
$\text{SNR}_A$. On the other hand, if $N$ is so small that the spectral codes are
undersampled, then $\gamma$ becomes reduced and accordingly so does $\text{SNR}_A$.

In our case, $\eta$ is approximately $0.06$ (corresponding to about $0.25$ per dimension), $\eta W$ is about $10^8$ photoelectrons/s, and $T \approx 1\,\rm s$. For a desired $\text{SNR}_A$ of, say, 10, we obtain $M$ roughly on the order of $100$. This is an order of magnitude estimate, roughly in accord with our results.

\section*{Discussion}

In summary, we have demonstrated a technique to perform imaging of
self-luminous objects through a single optical fiber. The technique is based
on encoding spatial information into spectral information that is immune to
fiber motion or bending. A key novelty of our technique is that encoding
is performed over the full spectral bandwidth of the system, meaning that
high throughput is maintained independent of the number of resolvable image
pixels (as opposed to wavelength-division encoding), facilitating the imaging of self-luminous objects. 

Encoding is performed by passive optical elements. In our demonstration we made use of low finesse FPEs, which exhibit no lateral structure and impart well-defined quasi-sinusoidal spectral
codes whose periods depend on illumination direction. Alternatively, SSEs
could impart much more complicated codes that are essentially random in
structure, as obtained, for example, from a large number of stacked
thin films of random thicknesses or refractive indices. SSE techniques based
on pseudo-random codes are akin to asynchronous CDMA techniques in wireless
communication. For our applications, these would render our image retrieval more immune to wavelength fading or color variations in our object spectrum.  It should be noted that spatio-spectral encoding based on
randomly scattering media has already been demonstrated \cite{dogariu,cao};
however in these cases the scattering was roughly isotropic rather than one-dimensional, which led to severe
beam spreading and ultimately limited throughput.

In principle, SSE's can be miniaturized and integrated within the optical fiber itself using photonic crystal or metamaterials techniques, or techniques similar to the
imprinting of fiber Bragg gratings. Such integrated encoders should exhibit structure almost exclusively in the axial direction, though some weak lateral structure may be required to break lateral degeneracies and enable 2D angular encoding (leading to slight loss of contrast in the spectral codes). Yet an alternative strategy to achieve 2D
encoding is to perform 1D encoding with a SSE, and combine this with a
physical translation or rotation of the optical fiber itself, to achieve 2D
imaging (similar to \cite{yelin, tearney2006}). This last alternative has the advantage that it can provide higher 2D image resolution, but has the disadvantage
that it requires moving parts. 

While our demonstration here has been limited to proof of concept, we anticipate that applications of our technique include fluorescence or chemiluminescence microendoscopy deep within tissue. The narrower bandwidths associated with such imaging would require finer structures in our spectral codes and concomitantly higher spectral resolution in our spectrograph. Nevertheless, in principle, even highly structured encoding can be obtained without loss of throughput and in an ultra-miniaturized geometry, leading to the attractive possibility of a microendoscopy device that causes minimal surgical damage.  Alternatively, our technique could be used for ultra-miniaturized imaging of self-luminous or white-light illuminated scenes for remote sensing.

\section*{Methods}

\noindent\textbf{Optical layout:} The principle of our layout is shown in Fig. \ref{Schematic} and described in more detail in Fig. \ref{SSE_setup}. An incoherent (e.g. self-luminous) object was simulated by sending collimated white light from a lamp through an intensity transmission spatial light modulator (Holoeye LC2002) with computer-controllable binary pixels. A lens ($f_\theta$) converted pixel positions into ray directions, which were then sent through the SSE and directed into an optical fiber ($200\,\rm \mu m$ diameter, 0.39NA). The spectrum at the proximal end of the fiber was recorded with a spectrograph (Horiba CP140-103) featuring a spectral resolution of $2.5\,\rm nm$ as limited by the core diameter of a secondary relay fiber.

\noindent\textbf{SSE construction:} Our SSE's were made of two low finesse FPEs tilted about $45^o$ relative to the optical axis in orthogonal directions relative to each other. The FPEs were made by evaporating thin layers of silver of thickness typically $17\,\rm nm$ onto standard microscope coverslips, manually pressing these together to obtain air gaps of several microns, and glueing with optical cement near the edges of the coverslips. Ideally the FPEs encoding orthogonal direction should exhibit very disparate free spectral ranges, however this was difficult to achieve in our case. The finesse of $\textrm{FPE}_1$ was about 5, measured at $670\,\rm nm$, with free spectral range of about $26\,\rm THz$, indicating that the air gap was approximately $6\,\rm\mu m$. The finesse of $\textrm{FPE}_2$ was about 3 at $670\,\rm nm$, with free spectral range $19\,\rm THz$ and air gap about $8\,\rm\mu m$.      

\noindent\textbf{Calibration and image retrieval:} To determine the SSE matrix $\textbf{M}$ prior to imaging, we cycled through each object pixel one by one and recorded the associated spectra (columns of $\textbf{M}$). Recordings were averaged over $20$ measurements with $800\,\rm ms$ spectrograph exposure times. Actual imaging was performed with single spectrograph recordings of $\rm{B}$ with $800\,\rm ms$ exposure times. Corresponding image pixels $A_m$ were reconstructed by least-squares fitting with a non-negativity prior, given by $\min_{\rm{A}}\Arrowvert \textbf{M}\rm{A-B} \Arrowvert^2$, where $A_m\ge0$ $\forall m$, as provided by the Matlab function \texttt{lsqnonneg}. We note that a baseline background spectrum obtained when all SLM pixels are off was systematically subtracted from all spectral measurements to correct for the limited on/off contrast (about 100:1) of our SLM.

\section*{Acknowledgments} We are grateful for financial support from the NIH and from the Boston University Photonics Center.

\section*{Author contributions} J.M. and R.B. conceived and designed the experiments. R.B. performed the experiments and analyzed the data. J.M. and R.B. wrote the paper.  

\section*{Competing financial interests}
The authors declare no competing financial interests.

\newpage

\newpage

\begin{figure}[h!]
\centerline{\includegraphics[width=12cm]{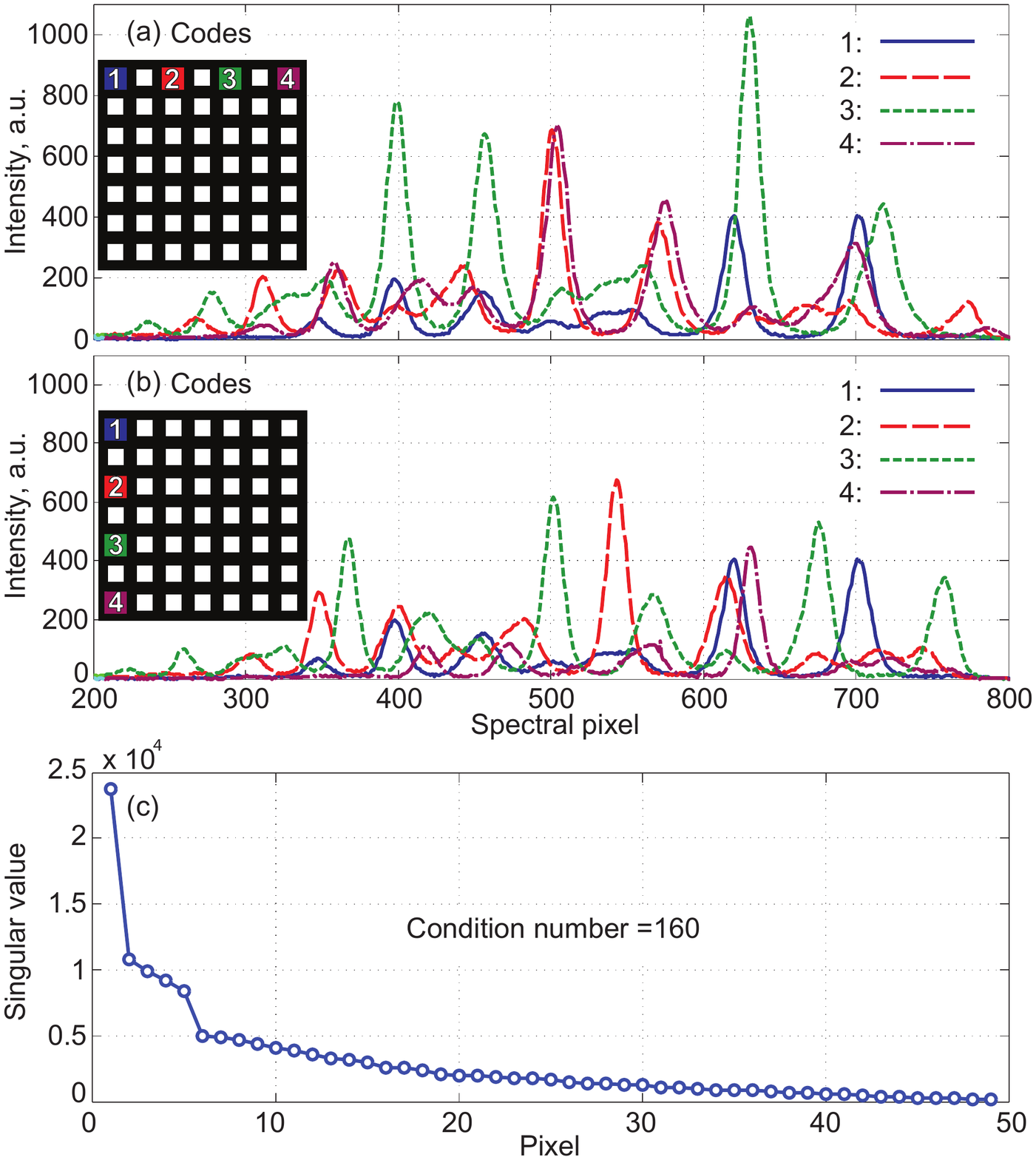}}
Supplementary Figure 1: {\bf Spectral codes used for object reconstruction in Fig.~3c.} Intensity profiles recorded by the spectrograph, where different plot colors represent spectra obtained when the object pixels (highlighted by same color in the insets) were sequentially illuminated along (a) a row, or (b) a column. (c) Singular values obtained from the singular value decomposition of the encoding matrix \textbf{M}. The condition number is defined as the ratio of the largest to smallest singular value. The spectral pixel marked $500$ corresponds to the wavelength $565\,\rm nm$; one spectral pixel spans about $0.5\,\rm nm$.   
\label{SVD_codes_suppl}
\end{figure}

\newpage

\begin{figure}[h!]
\centerline{\includegraphics[width=12cm]{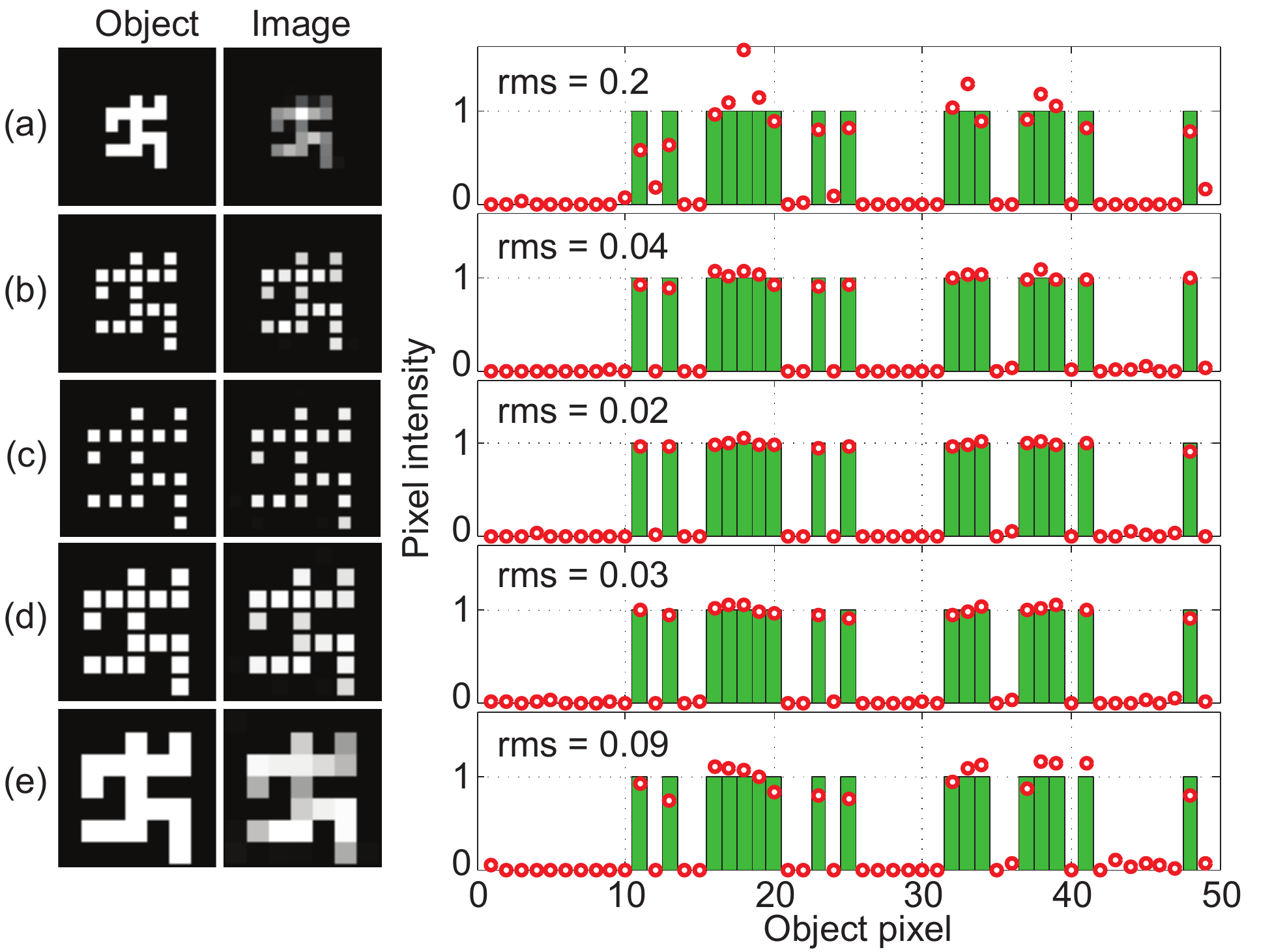}}
Supplementary Figure 2: {\bf Influence of pixel size and sparcity on SSE imaging.} Spatially incoherent, white-light objects in the shape of a running man are sent through a SSE and launched into a fiber (see Methods). The reconstructed images and pixel values, along with the associated rms errors, are displayed for different pixel layouts. In rows (a)-(c), the pixel sizes (powers) are the same, but their separation is increased, leading to better reconstruction (lower rms error). In rows (c)-(e) the pixel separations are the same, but their size is increased until the pixels are juxtaposed. The corresponding increase in pixel power is counterbalanced by a decrease in the contrast of associated spectral codes (due to increased angular diversity per pixel), leading to a net decrease in reconstruction quality (higher rms).
\label{RM}
\end{figure}

\newpage

\begin{figure}[h!]
\centerline{\includegraphics[width=14cm]{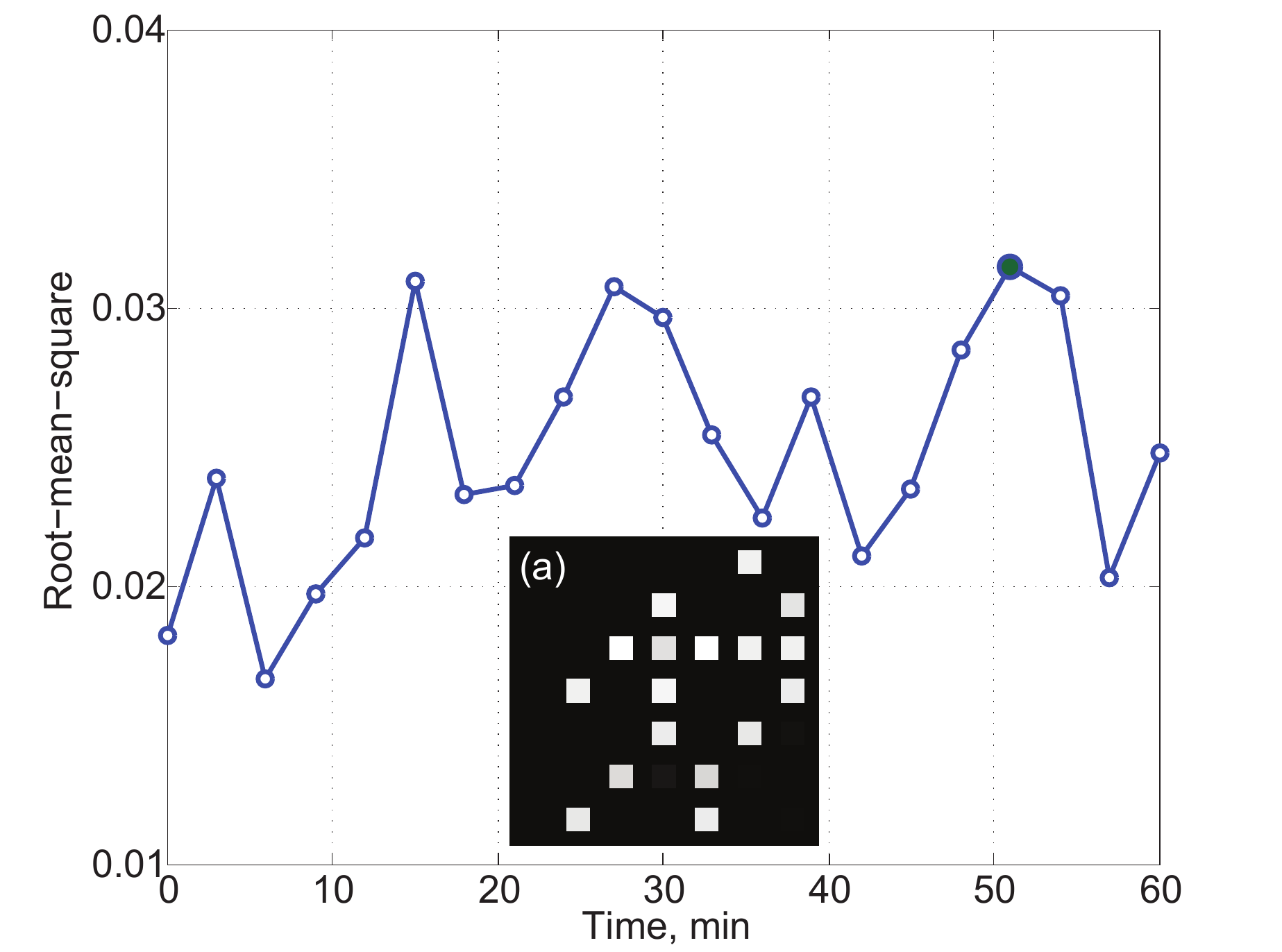}}
Supplementary Figure 3: {\bf Demonstration of stability of image reconstruction.} Evolution of the image reconstruction quality, quantified by the root-mean-square (rms) error, as a function of time. The coding matrix measured prior to zero time was used to reconstruct an image in the shape of an archer from spectra acquired at different elapsed times. The inset illustrates the image reconstructed at 51 min.  
\label{RMS_time}
\end{figure}

\end{document}